\documentclass[LATO1COL]{WileyNJD-v2}
\usepackage{moreverb}
\historydates{}
\doiheadtext{} 
\articletype{}
\gdef\printjnlcitation{}%

\newcommand\BibTeX{{\rmfamily B\kern-.05em \textsc{i\kern-.025em b}\kern-.08em
T\kern-.1667em\lower.7ex\hbox{E}\kern-.125emX}}

\usepackage[utf8]{inputenc}
\usepackage{glossaries}
\newacronym{gpc}{gPC}{generalized polynomial chaos}
\newacronym{pdf}{PDF}{probability density function}
\newacronym[longplural={quantities of interest}]{qoi}{QoI}{quantity of interest}
\newacronym{rv}{RV}{random variable}
\newacronym{fe}{FE}{finite element}
\newacronym{pec}{PEC}{perfect electric conductor}
\newacronym{uq}{UQ}{uncertainty quantification}
\newacronym{mim}{MIM}{metal-insulator-metal}
\makeglossaries
\usepackage{subcaption}
\usepackage{siunitx}
\usepackage{amsfonts}
\usepackage{mathtools}

\begin{document}

\title{Conformally Mapped Polynomial Chaos Expansions for Maxwell's Source Problem with Random Input Data\protect} 
\author[1,2]{Niklas Georg*}

\author[1]{Ulrich Römer}

\authormark{Niklas Georg and Ulrich Römer}

\address[1]{\orgdiv{Institut für Dynamik und Schwingungen (IDS)}, \orgname{Technische Universität Braunschweig}, \orgaddress{\country{Germany}}}

\address[2]{\orgdiv{Centre for Computational Engineering (CCE) and Institute for Accelerator Science and Electromagnetic Fields (TEMF)}, \orgname{Technische Universität Darmstadt}, \orgaddress{ \country{Germany}}}

\corres{*Niklas Georg, Schleinitzstraße 20, \\38106 Braunschweig, Germany\\ \email{n.georg@tu-braunschweig.de}}

\abstract[Abstract]{Generalized Polynomial Chaos (gPC) expansions are well established for forward uncertainty propagation in many application areas. Although the associated computational effort may be reduced in comparison to Monte Carlo techniques, for instance, further convergence acceleration may be important to tackle problems with high parametric sensitivities. In this work, we propose the use of conformal maps to construct a transformed gPC basis, in order to enhance the convergence order. The proposed basis still features orthogonality properties and hence, facilitates the computation of many statistical properties such as sensitivities and moments. The corresponding surrogate models are computed by pseudo-spectral projection using mapped quadrature rules, which leads to an improved cost accuracy ratio. We apply the methodology to Maxwell's source problem with random input data. In particular, numerical results for a parametric finite element model of an optical grating coupler are given.}


\maketitle

\section{Introduction}\label{sec1}
Due to recent developments in \gls{uq} \cite{xiu2009}, studying random parameter variations within the numerical simulation of fields and waves comes into reach. The present study is motivated from the design of optical components and plasmonic structures, where relatively large variabilities of nano-scale geometrical parameters can be observed, see, e.g. \cite{preiner2008}.
In this work, we focus on the forward problem, i.e., the propagation of uncertainties from the model inputs to the outputs, in order to compute statistics and sensitivities for physical \glspl{qoi}. We rely on surrogate modeling \cite{lemaitre2010} to reduce the computational complexity of sampling the underlying \gls{fe} Maxwell solver. Although motivated from a forward model perspective, the surrogate construction could equally be used in an inverse problem context. Examples of surrogate modeling in electromagnetics can be found for instance in \cite{austin2013, Georg_2019aa, Corno2019} where microwave circuits and accelerator cavities are considered. 

\Gls{gpc} expansions \cite{xiu2002} are powerful tools for forward uncertainty propagation. They are based on an orthogonal polynomial basis with respect to the underlying probability distribution of the input parameters, to achieve good convergence properties. However, applying \gls{gpc} may still be challenging, the computational cost to handle large parameter uncertainties and parametric sensitivities for instance may be quite high. To remedy this issue, conformal maps can be utilized in order to improve the convergence of polynomial-based methods. The acceleration of quadrature methods by the use of conformal maps, has been considered in \cite{hale2008, trefethen2013, jantsch2018sparse}. In \cite{georg2018}, conformal maps were combined with a stochastic collocation method, indicating significant gains in the accuracy of the corresponding surrogate model. In this work, we propose a new orthogonal basis by combining \gls{gpc} and conformal maps. 
We note that, the proposed basis is constructed such that it fulfills the same orthogonality properties as \gls{gpc}. Accordingly, advantages of \gls{gpc} methods are preserved, e.g., stochastic moments and Sobol coefficients can be directly computed from the expansion coefficients. It should also be noted that various approaches employing Polynomial Chaos expansions with basis rotation have been reported recently, see \cite{tsilifis2017reduced,papaioannou2019pls}. Although, these works equally rely on mapped Polynomial Chaos approximations, the transformations are linear (affine) and not based on conformal mappings. Also, the emphasis there is on high dimensional approximation instead of convergence acceleration.

\begin{figure}	
\begin{subfigure}[b]{.6\textwidth}
\centering
\includegraphics{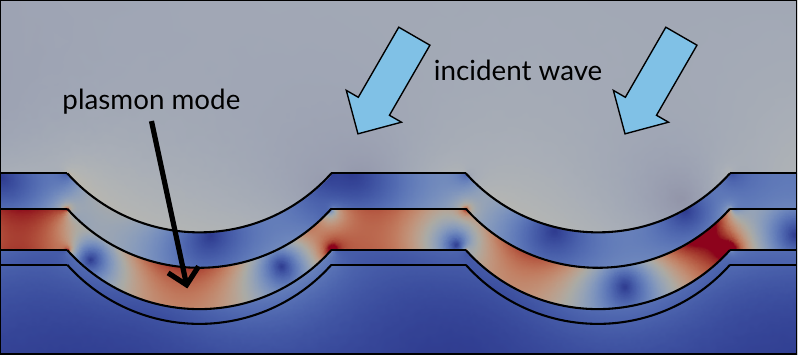}
\caption{Optical coupling into MIM plasmon modes \cite{preiner2008}.}
\label{fig:mimPlasmonMode}
\end{subfigure}
\begin{subfigure}[b]{.38\textwidth}
\centering
\includegraphics{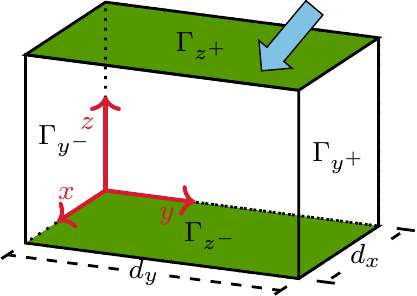}
\caption{Sketch of considered unit cell corresponding to the computational domain $\ensuremath{D}$. The blue arrow illustrates the incident wavevector $\mathbf k^\mathrm{inc}$.}
\label{fig:unit_cell}
\end{subfigure}
\caption{Scattering of periodic structure excited by an incident plane wave.}
\end{figure}

The proposed numerical scheme is applied to quantify uncertainties via surrogate models for Maxwell's equations in the frequency domain. In particular, we consider the source problem on periodic domains with a plane wave excitation and uncertainties in the material interface geometry. Such model equations can describe, for instance, the coupling into \gls{mim} plasmon modes with subwavelength diffraction gratings, which is illustrated in Fig.~\ref{fig:mimPlasmonMode}. Although illustrated by means of this particular application example, we note that the employed \gls{uq} methodologies apply in a much broader context.

This paper is structured as follows: Section \ref{sec:maxwell} contains a brief description of Maxwell's source problem. The uncertainty quantification part can be found in Section \ref{sec:uq}, where we briefly recall standard \gls{gpc} before discussing the proposed extension based on conformal mappings. Section \ref{sec:application} reports numerical results for an analytical RLC circuit and the aforementioned optical grating coupler, before conclusions are drawn.  

\section{Maxwell's Source Problem}
\label{sec:maxwell}
We consider Maxwell's source problem for periodic structures excited by an incident plane wave. For further details on this subject, we refer to \cite{georg2018, jin2015}. We start with the time-harmonic curl-curl equation 
\begin{equation}
\nabla \times \left(\mu_{\mathrm{r}}^{-1} \nabla \times \ensuremath{\mathbf{E}}\right)- \omega^2 \varepsilon\mu_0 \ensuremath{\mathbf{E}} = 0   \qquad \text{in }\ensuremath{D}, \label{eq:curlcurl}
\end{equation}
for the electric field phasor $\ensuremath{\mathbf{E}}$ in the computational domain \ensuremath{D}, where $\omega$ denotes the angular frequency, $\varepsilon$ the complex permittivity and $\mu_{\mathrm r}, \mu_0$ denote the relative and vacuum permeability, respectively. Note that \eqref{eq:curlcurl} assumes absence of charges and source currents in \ensuremath{D}. Based on Floquet's theorem \cite[Chapter 13]{jin2015}, the computational domain $\ensuremath{D}$ can be reduced to a unit cell of the periodic structure, as we assume a periodic excitation. Such a unit cell is depicted in Fig.~\ref{fig:unit_cell}. Due to the oblique angle of the incident wave, the excitation has a different periodicity than the geometry and, hence, periodic phase-shift boundary conditions need to be imposed on the respective boundaries. To truncate the structure in the non-periodic direction, a Floquet absorbing boundary condition and a \gls{pec} boundary condition are applied. This leads to the boundary value problem
\begin{align}
\nabla \times \left(\mu_{\mathrm{r}}^{-1} \nabla \times \ensuremath{\mathbf{E}}\right)- \omega^2 \varepsilon\mu_0 \ensuremath{\mathbf{E}} &= 0  &&\text{in }D\\
\ensuremath{\mathbf{E}}\vert_{\ensuremath{{\Gamma_{x^+}}}} e^{j\ensuremath{k^{\mathrm{inc}}_x} \ensuremath{{d_x}}}&= \ensuremath{\mathbf{E}}\vert_{\ensuremath{{\Gamma_{x^-}}}}  \ \ &&\text{on}~{\ensuremath{{\Gamma_{x^+}}} \cup \ensuremath{{\Gamma_{x^-}}}}\\
\ensuremath{\mathbf{E}}\vert_{\ensuremath{{\Gamma_{y^+}}}} e^{j\ensuremath{k^{\mathrm{inc}}_y} \ensuremath{d_y}}&= \ensuremath{\mathbf{E}}\vert_{\ensuremath{{\Gamma_{y^-}}}}  \ \ &&\text{on}~{\ensuremath{{\Gamma_{y^+}}}\cup \ensuremath{{\Gamma_{y^-}}}}\\
\ensuremath{\mathbf{n}} \times \ensuremath{\mathbf{E}} &= 0  && \text{on }\ensuremath{{\Gamma_{z^-}}} \\
(\mu_\mathrm{r}^{-1} \nabla \times \ensuremath{\mathbf{E}})\times \ensuremath{\mathbf{n}} + \mathcal{F}(\ensuremath{\mathbf{E}} ) &= \mathcal{G}(\ensuremath{\mathbf{E}}^{\mathrm{inc}}) \ \ &&\text{on } \ensuremath{{\Gamma_{z^+}}},
\end{align}
where we refer to \cite[Appendix A]{georg2018} for a derivation and definition of the functionals $\mathcal F(\cdot), \mathcal G(\cdot)$.

We assume in the following that the complex permittivity $\varepsilon$ depends smoothly on a parameter vector $\ensuremath{\mathbf{y}}\in \Xi \subset \mathbb R^N$. These parameters can then be used to model variations in the refractive indices or extinction coefficients of the (different) materials in $D$, as well as changes in the geometry of the material interfaces inside the domain $D$.  
Following a standard Galerkin procedure, cf. \cite{georg2018}, we then obtain a \gls{fe} model in the form 
\begin{equation}
\label{eq:pde_weak}
\text{find }\ensuremath{\mathbf{e}}(\ensuremath{\mathbf{y}}) \in  V \text{ s.t.} \quad  a_\ensuremath{\mathbf{y}}(\ensuremath{\mathbf{e}}(\ensuremath{\mathbf{y}}),\ensuremath{\mathbf{v}})  =  l_\ensuremath{\mathbf{y}}(\ensuremath{\mathbf{v}}) \quad \forall \ensuremath{\mathbf{v}} \in  V, 
\end{equation}
where $a_{\mathbf y}(\cdot, \cdot)$ is a continuous sesquilinear form, $l_{\mathbf y}(\cdot)$ is a continuous (anti)linear form and $V$ denotes a discrete subspace of ${\ensuremath{\mathbf{H}\left(\text{curl};\ensuremath{D}\right)}}$ \cite{monk2003}, enforcing periodic phase-shift conditions on the traces at the periodic boundaries and homogeneous Dirichlet conditions at $\Gamma_{z^-}$. To achieve a curl-conforming discretization of \eqref{eq:pde_weak}, we employ N{\'e}d{\'e}lec's elements of the first kind \cite{nedelec1980} and 2nd order on a tetrahedral mesh of $\ensuremath{D}$. As \gls{qoi} we consider the fundamental reflection coefficient $\mathcal Q(\mathbf e(\mathbf y))$, i.e. a scattering parameter, which can be computed as an affine-linear functional of the electric field $\ensuremath{\mathbf{e}}$ in post-processing \cite{georg2018}. For brevity, we replace $\mathcal Q(\mathbf e(\mathbf y))$ by $\mathcal Q(\mathbf y)$ in the following.

\section{Uncertainty Quantification}
\label{sec:uq}
To account for uncertainty, we model the input parameters $\mathbf y$ as independent \glspl{rv} with joint probability density function $\rho$ and image set $\Xi\subset \mathbb R^N$, where we assume in this section for brevity of notation that $\Xi$ is given as the hypercube $[-1,1]^N$.  Note that different image sets $\Xi$ or stochastic dependence could also be considered, e.g. by a Rosenblatt transformation \cite{lebrun2009rosenblatt}. Additionally, we assume that the map $\mathcal Q: \Xi \rightarrow \mathbb C$ is holomorphic. Note that this assumption can often be justified for boundary value problems with random influences, see, e.g., \cite{hiptmair2018}. Holomorphy of the solution of Maxwell's source problem with respect to general shape parametrizations was established in \cite{aylwin2019}.

As discussed in the following, in this work we propose a method for surrogate modeling, where the basis functions are mapped polynomials based on \gls{gpc} \cite{xiu2002} combined with a conformal mapping. To compute the corresponding coefficients we rely on pseudo-spectral projection based on mapped quadrature rules \cite{hale2008}. 

\subsection{Generalized Polynomial Chaos}
For convenience of the reader, we briefly recall the standard polynomial chaos expansions, going back to Wiener \cite{wiener1938}. Considering Gaussian random variables, any $\mathcal Q(\mathbf y)$ with bounded variance, can be accurately represented using Hermite polynomials as basis functions. Employing the Askey-scheme \cite{xiu2002}, for different probability distributions $\rho$, basis functions $\Psi_m:\Xi \rightarrow \mathbb R$ which are orthonormal w.r.t. the probability density $\rho$, i.e.,  
\begin{equation}
\mathbb E[\Psi_i\Psi_j] := \int_\Xi \Psi_i(\mathbf y)\Psi_j(\mathbf y) \rho(\mathbf y) \,\mathrm d\mathbf y =\delta_{ij}, \label{eq:orth}
\end{equation}
can be obtained. We note that \gls{gpc} can also be constructed for arbitrary densities $\rho$ \cite{soize2004}. The \gls{gpc} approximation is then given as 
\begin{equation}
\label{eq:mtermsapprox}
\mathcal Q_M^{\mathrm{PC}}\left(\mathbf{y}\right) 
= \sum_{m=0}^{M} s_m \Psi_m\left(\mathbf{y}\right),
\end{equation}
where the $s_m \in \mathbb{C}$ denote the \gls{gpc} coefficients. 
In practice, in order to obtain a computable expression, the sum in \eqref{eq:mtermsapprox} has to be truncated to $M<\infty$ and limited polynomial degrees are considered. 
The coefficients $s_m$ can then be determined in various ways, e.g. by regression or stochastic collocation, see \cite{xiu2010} for an overview. Here we consider projection, i.e.,
\begin{equation}
s_m = \mathbb E[\mathcal Q\Psi_m] = \int_\Xi \mathcal Q(\mathbf y) \Psi_m(\mathbf y) \rho(\mathbf y)\,\mathrm{d}\mathbf y. \label{eq:gpc_proj}
\end{equation}
The integral in \eqref{eq:gpc_proj} is usually not readily computable and is hence often approximated by numerical quadrature. Due to orthogonality of the basis, stochastic moments as well as variance-based sensitivity indices can then be calculated directly from the coefficients $s_m$ without further approximations, see \cite{xiu2010}.
These methods show spectral convergence, e.g.,  in the norm $||u||_{L^2_\rho}:=\sqrt{\mathbb E[u^2]}$  \cite{xiu2002}. In particular, if the map $\mathbf y\mapsto \mathcal Q(\mathbf y)$ is analytic, exponential convergence can be expected, as discussed in the following. Note that, for simplicity, we first consider the univariate case, i.e., $N=1$, while generalizations to the multivariate case $N>1$ will be discussed later.

We assume that $\mathcal Q_\mathrm{1D}:[-1,1]\rightarrow\mathbb C$ can be analytically extended onto an open Bernstein ellipse $E_r\subset \mathbb C$. A Bernstein ellipse $E_r$ is an ellipse with foci at $\pm1$ and the size $r$ is given by the sum of the length of semi-major and semi-minor axis. This is illustrated in Fig.~\ref{fig:Bernstein}. Following \cite{trefethen2013}, the error of the polynomial best approximation $\mathcal Q_{M}^\mathrm{PC^*}$ with degree $M$ can be estimated as 
\begin{align}
\|\mathcal Q_\mathrm{1D}-\mathcal Q_{M}^\mathrm{PC^*}\|_\infty &\le \frac {C_\text{B} r^{-M}}{r-1}, \label{eq:conv_est}
\end{align}
where $\|\cdot\|_\infty$ denotes the supremum-norm on $[-1,1]$  and the constant $C_\text{B}>0$ depends on the uniform bound of $\mathcal Q_\mathrm{1D}$ in $E_r$. Note that convergence in the supremum-norm implies convergence in the $||\cdot||_{L^2_\rho}$ norm as well, as
\begin{equation}
||\mathcal Q_\mathrm{1D}-\mathcal Q_{M}^\mathrm{PC^*}||_{L^2_\rho} =\Bigl( \int_{[-1,1]} \bigl(\mathcal Q_\mathrm{1D}-\mathcal Q_{M}^\mathrm{PC^*}\bigr)^2\rho_{\mathrm{1D}} \,\mathrm d y\Bigr)^{\frac 1 2} \le \|\sqrt{\rho_{\mathrm{1D}}}\|_\infty \,\|\mathcal Q_\mathrm{1D}-\mathcal Q_{M}^\mathrm{PC^*}\|_\infty  \Bigl(\int_{[-1,1]} 1\,\mathrm dy \Bigr)^{\frac 1 2}=\sqrt{2} \,\|\sqrt{\rho_{\mathrm{1D}}}\|_\infty \,\|\mathcal Q_\mathrm{1D}-\mathcal Q_{M}^\mathrm{PC^*}\|_\infty.  \end{equation}
We further note that the additional aliasing error introduced by the discrete projection does not harm the convergence order for well-resolved smooth function, cf.~\cite[Chapter 3.6]{xiu2002}.

\subsection{Conformally Mapped Generalized Polynomial Chaos}
\begin{figure}
\begin{subfigure}[b]{0.45\textwidth}
\includegraphics{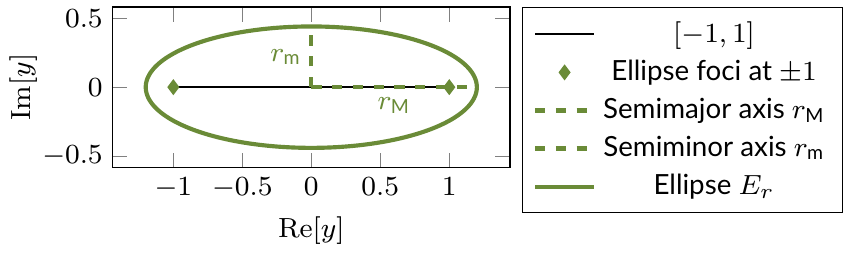}
\caption{Bernstein ellipse $E_r$ of size $r=r_{\text M}+r_\text{m}$.}
\label{fig:Bernstein}
\end{subfigure}\hfill
\begin{subfigure}[b]{0.5\textwidth}
\centering 
\includegraphics{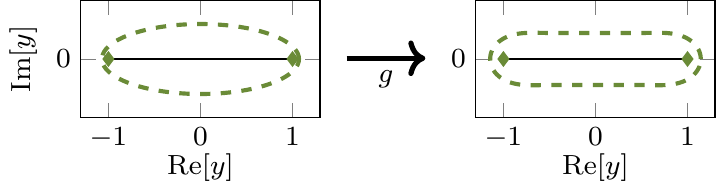}
\vspace{.5em}
\caption{Conformal map of a Bernstein ellipse $E_r$ (left) to \textit{straighter} region $g(E_r)$.}
\label{fig:mapping}
\end{subfigure}
\caption{Illustration of conformal mapping approach.}
\end{figure}

Equation \eqref{eq:conv_est} shows that the convergence is connected to the region of analyticity, in particular the convergence order $r$ depends on the size of the largest Bernstein ellipse not containing any poles of the continuation of $\mathcal Q_\mathrm{1D}$ (in the complex plane). However, established procedures \cite{babuska2007} inferring the regularity of parametric problems based on a sensitivity analysis, do not lead to elliptical regions, but rather prove analyticity in an $\epsilon$-neighborhood of the unit interval. In this case, a conformal map $g$ can be employed, which maps Bernstein ellipses to \textit{straighter} regions and thus, enlarges the domain of analyticity, as illustrated in Fig.~\ref{fig:mapping}. 
To this end, there are various mappings which could be employed, cf. \cite{hale2009}. Here, we focus for simplicity on the so-called $9$-th order sausage mapping 
\begin{equation}g(s) = \frac 1 {53089} (40320s+6720s^3+3024s^5+1800s^7 +1225s^9)\label{eq:sausage_map}\end{equation}
introduced in \cite{hale2008}, which represents a normalized Taylor approximation of the inverse sine function. Note that $g$ maps the unit interval to itself, i.e., 
\begin{equation}
g([-1,1])=[-1,1] ~\text{and}~g(\pm 1)=\pm1.\label{eq:unit_interval}
\end{equation}

Conformal maps were employed in \cite{hale2008} to derive new numerical quadrature formulas, and have also recently been considered in the context of stochastic collocation methods \cite{jantsch2018sparse, georg2018}. 
In this work, we address the combination of conformal maps and polynomial chaos expansions. Based on the assumption that $h:=\mathcal Q_\mathrm{1D}\circ g$ has a larger Bernstein ellipse than $\mathcal Q_\mathrm{1D}$, and is hence better suited to be approximated with polynomials, we propose a new orthogonal basis
\begin{equation}
\Phi_m \coloneqq \tilde \Psi_m\circ g^{-1},\quad m=0,\ldots,M \label{eq:map_basis}
\end{equation}
where $\tilde \Psi_m$ are orthonormal polynomials w.r.t. the transformed density 
\begin{equation}
\tilde \rho_{\mathrm{1D}} (s) := g'(s) \rho_{\mathrm{1D}}(g(s)).
\label{eq:transformed_pdf}
\end{equation}
We emphasize that 	$\{\Phi_m\}_{m=0}^M$ forms an orthonormal basis w.r.t. the input probability distribution $\rho$. This can be shown by a change of variables $y=g(s)$
\begin{align}
\mathbb E[\Phi_i \Phi_j] &= \int_{-1}^1 (\tilde \Psi_i\circ g^{-1})(y)(\tilde \Psi_j\circ g^{-1})(y) \rho_{\mathrm{1D}}(y)\,\mathrm{d}y \\
&=\int_{-1}^1 \tilde \Psi_i(s) \tilde \Psi_j(s) \underbrace{\rho_{\mathrm{1D}}(g(s))g'(s)}_{\tilde \rho_{\mathrm{1D}}(s)}\,\mathrm{d}s= \delta_{ij}, 
\end{align}
where the last line holds by construction of the polynomials $\tilde \Psi_m$. 
Due to the orthogonality, the corresponding coefficients $s_m$ of the mapped approximation 
\begin{equation}
{Q}_M(y) = \sum_{m=0}^M s_m \Phi_m(y) \label{eq:map_approx}
\end{equation}
can then be determined by the projection
\begin{equation}
s_m = \mathbb E[\Phi_m Q_\mathrm{1D}] = \int_{-1}^1  \Phi_m(y) Q_\mathrm{1D}(y) \rho_{\mathrm{1D}}(y)\,\mathrm{d}y. \label{eq:mapped_proj}
\end{equation} Note that, by abuse of notation, we use the same symbol $s_m$ for the \gls{gpc} coefficients and the mapped \gls{gpc} coefficients. 
The mapped polynomial best approximation $ \mathcal Q_M^*$ converges as 
\begin{align}
\|\mathcal Q_\mathrm{1D}- {\mathcal Q}_M^*\|_\infty &= \|\mathcal Q_\mathrm{1D}\circ g-{\mathcal Q}_M^*\circ g\|_\infty =\|(h-h_M^{\mathrm{PC}^*}) \|_\infty \le \frac{\tilde C_B \tilde r^{-M}}{\tilde r-1},
\end{align}
where $h_M^{\mathrm{PC}^*}$ denotes the polynomial best approximation of $h$ and $\tilde r$ the size of a Bernstein ellipse $E_{\tilde r}$ on which an analytic continuation of $h$ exists. 

In particular, the convergence order of the mapped approximation $\mathcal Q_M$ is given by the size of the largest Bernstein ellipse $E_{\tilde r_{\mathrm{max}}}$ which is fully mapped into the region of analyticity of $\mathcal Q_\mathrm{1D}(y)$.  Note that $\tilde r_{\mathrm{max}}>r_{\mathrm{max}}$ for any positive $\epsilon<0.75$ \cite{georg2018}, and hence, a convergence improvement is to be expected in those cases, i.e., for functions analytic in such $\epsilon$-neighborhoods. It should be mentioned nevertheless that this procedure does not always yield improved convergence rates. One can easily imagine poles located such that a Bernstein ellipse may lead to a larger region of analyticity than a strip-like geometry. In the examples considered in this work, however, convergence acceleration could indeed be obtained.

To numerically compute \eqref{eq:mapped_proj}, we derive mapped quadrature rules, cf. \cite{hale2008, hale2009}. As pointed out in \cite{trefethen2013} for instance, Gaussian quadrature is derived from polynomial approximations and, hence, the convergence order also depends on the size of the Bernstein ellipse corresponding to the regularity of the integrand, see e.g. \cite[Theorem 1]{hale2008}. Therefore, relying again the assumption that $\mathcal Q_\mathrm{1D}\circ g$ has a larger Bernstein ellipse, we apply a change of variables $y=g(s)$ in \eqref{eq:mapped_proj}
\begin{equation}
s_m = \mathbb E[\Phi_m \mathcal Q_\mathrm{1D}] = \int_{-1}^1  \Phi_m(y) \mathcal Q_\mathrm{1D}(y) \rho_{\mathrm{1D}}(y)\,\mathrm{d}y =
\int_{-1}^1  \Phi_m(g(s)) \mathcal Q_\mathrm{1D}(g(s)) \underbrace{\rho_{\mathrm{1D}}(g(s)) g'(s)}_{\tilde \rho_{\mathrm{1D}}}\,\mathrm{d}s. \label{eq:mapped_proj_trans}
\end{equation}
The mapped quadrature scheme is then obtained by application of Gaussian quadrature w.r.t. the transformed density $\tilde \rho_{\mathrm{1D}}$, i.e. quadrature nodes $\{\tilde y^{(i)}\}_{i=0}^{M_\mathrm{quad}}$ and correspondings weights $\{\tilde w^{(i)}\}_{i=0}^{M_\mathrm{quad}}$, to the transformed integrand in \eqref{eq:mapped_proj_trans} 
\begin{equation}
s_m \approx \sum_{i=0}^{M_\text{quad}}  \Phi_m(g(\tilde y^{(i)}))  \mathcal Q_\mathrm{1D}(g( \tilde y^{(i)})) \tilde w^{(i)} = \sum_{i=0}^{M_\text{quad}} \Phi_m(\hat y^{(i)})  \mathcal Q_\mathrm{1D}(\hat y^{(i)})\hat w^{(i)}. 
\end{equation}
Note that the mapped quadrature nodes are obtained as $\hat y^{(i)} := g(\tilde y^{(i)})$, while the mapped weights are given as $\hat w^{(i)} := \tilde w^{(i)}$. Due to \eqref{eq:unit_interval}, it is ensured that the mapped quadrature nodes $\hat y^{(i)}$ do not require the evaluation of the analytic continuation of $\mathcal Q_\mathrm{1D}$ in the complex plane, which is, in practice, not always possible. A convergence improvement is expected based on the assumption that the transformed integrand in \eqref{eq:mapped_proj_trans} has a larger Bernstein ellipse. For further details on mapped quadrature schemes, we refer to \cite{hale2008}. However, we note the (minor) differences that in this work we employ Gaussian quadrature w.r.t. the transformed density $\tilde \rho_{\mathrm{1D}}$ to derive the mapped quadrature scheme, while \cite{hale2008} only considers unweighted Gaussian quadrature and, thereby, takes $g'(s)$ as part of the integrand (instead of the weight).

We proceed with a discussion of the multivariate case $N>1$. To this end, we introduce the multivariate mapping $\mathbf g(\mathbf s) = [g_1(s_1), \ldots, g_N(s_N)]$. In this work, we employ, for simplicity, the same mapping  \eqref{eq:sausage_map} for all parameters, i.e. $g_1=\ldots=g_N=g$. However, different choices would be possible as well. We also note that, for the trivial mapping $\mathbf g_\mathrm{triv}: \mathbf s\mapsto \mathbf s$  standard polynomial chaos expansions would be recovered. For each parameter $y_i$ with univariate \gls{pdf} $\rho_i$, we define the transformed \gls{pdf} $\tilde \rho_i(y_i) := \rho_i(g_i(y_i)) g_i'(y_i)$. The corresponding transformed joint \gls{pdf} is then given by $\tilde \rho(\mathbf y) = \tilde \rho_1(y_1) \ldots \tilde \rho_N(y_N)$. In the following, we denote by $\{\tilde \Psi_{\mathbf{m}}\}_{\mathbf m}$ an orthonormal polynomial basis w.r.t. to the transformed density $\tilde \rho$, i.e. \begin{equation}
\mathbb E_{\tilde\rho}[\tilde \Psi_{\mathbf i}\tilde \Psi_{\mathbf j}]:=\int_{\Xi}\tilde \Psi_{\mathbf i}(\mathbf y)\tilde \Psi_{\mathbf j}(\mathbf y)\tilde\rho(\mathbf y) \,\mathrm d\mathbf y = \delta_{i_1j_1}\ldots\delta_{i_Nj_N}, \label{eq:orth_ND}
\end{equation}
where we introduced the multi-index $\mathbf m = (m_1,\ldots, m_N)$ holding the univariate polynomial degrees, such that $\tilde \Psi_{\mathbf m}$ is a tensor-product polynomial of order $m_j$ in dimension $j=1,\ldots, N$. The respective mapped polynomials are then obtained as 
\begin{equation}\Phi_{\mathbf m}(\mathbf y) :=  (\tilde \Psi_{\mathbf m} \circ \mathbf g^{-1})(\mathbf y).
\end{equation}
The coefficients of the multivariate mapped approximation 
\begin{equation}
{\mathcal Q}_p(\mathbf y) := \sum_{\| \mathbf m\|_\infty \le p} s_{\mathbf m} \Phi_{\mathbf m}(\mathbf y), \label{eq:map_approx_ND}
\end{equation}
where we consider for simplicity a tensor-product construction of maximum degree $p$, can then again be obtained by projection 
\begin{equation}
s_{\mathbf m} = \mathbb E[ \Phi_{\mathbf m} \mathcal Q] = \int_\Xi \Phi_{\mathbf m}(\mathbf y) \mathcal Q(\mathbf y) \rho(\mathbf y)\,\mathrm{d}\mathbf y. \label{eq:multivariate_projection}
\end{equation}
To evaluate the multi-dimensional integral in \eqref{eq:multivariate_projection}, we employ mapped Gaussian quadrature. In this case the mapped nodes and weights are given by $\hat{\mathbf y}^{(i)} := \mathbf g(\tilde{\ensuremath{\mathbf{y}}}^{(i)})$ and $\hat w^{(i)}:=\tilde w^{(i)}$, respectively,  where, in turn, $\tilde{\ensuremath{\mathbf{y}}}^{(i)}$ and $\tilde w^{(i)}$ are the nodes and weights of a Gaussian quadrature w.r.t. $\tilde \rho$.

Finally, we emphasize that, since the mapped representation \eqref{eq:map_approx_ND} uses an orthogonal basis, the coefficients $s_{\mathbf m}$ can be used to directly compute stochastic moments as well as variance-based sensitivity indices. For instance, the mean value is given by 
\begin{equation}
\mathbb E[{\mathcal Q}_p] = \int_{\Xi }  \Bigl(\sum_{\| \mathbf m\|_\infty \le p} s_{\mathbf m} \Phi_{\mathbf m}(\mathbf y)\Bigr) \rho(\mathbf y)\,\mathrm d\mathbf y =  s_{\mathbf 0},
\end{equation}
where we employed, that the mapped basis function $\Phi_{\mathbf 0}$ is constant on $\Xi$, as well as the orthonormality condition \eqref{eq:orth_ND}. Accordingly the variance is given by 
\begin{equation}
\mathbb V[{\mathcal Q}_p]=\mathbb E[{\mathcal Q}_p^2] -\mathbb E[{\mathcal Q}_p]^2 = \sum_{0<\|\mathbf m\|_\infty\le p} s_{\mathbf m}^2. 
\end{equation}
Additionally, Sobol sensitivity indices \cite{Sobol2001}, based on a decomposition of the variance, can also be directly derived from the coefficients. Regarding the estimation of Sobol indices, we will focus on the so-called main-effect (1st order) and total-effect (total order) indices.
We define the multi-index sets $\Lambda_n^{\text{main}}, \Lambda_n^{\text{total}} \subset \Lambda^\text{TP}_{p}:=\{\mathbf m\,|\, 0\le \|\mathbf m \|_\infty \le p\}$, $n=1,2,\dots,N$, such that 
\begin{align}
\Lambda_n^\text{main} &= \{\mathbf{m} \in \Lambda_{p}^\text{TP} \; : \; m_n \neq 0 \hspace{0.5em} \text{and} \hspace{0.5em} m_j = 0, n \neq j\},\\	\Lambda_{n}^{\text{total}} &= \{\mathbf{m} \in \Lambda^\text{TP}_{p} \; : \; m_n \neq 0\}.
\end{align}
We then define the partial variances $\mathbb V_n^{\text{main}}\left[{\mathcal Q}_p\right]$ and $\mathbb V_n^{\text{total}}\left[{\mathcal Q}_p\right]$, such that
\begin{align}
\mathbb V_n^{\text{main}}\left[{\mathcal Q}_p\right] = \sum_{\mathbf{m}  \in \Lambda_n^{\text{main}}} s_{\mathbf{m}}^2, &&
\mathbb V_n^{\text{total}}\left[{\mathcal Q}_p\right]= \sum_{\mathbf{m}  \in \Lambda_n^{\text{total}}} s_{\mathbf{m}}^2.
\end{align}
Then, the main-effect and total-effect Sobol indices, $ S_n^{\text{main}}$ and $S_n^{\text{total}}$, respectively, are given as
\begin{align}
S_n^{\text{main}}[{\mathcal Q}_p] = \frac{\mathbb V_n^{\text{main}}\left[{\mathcal Q}_p\right]}{\mathbb V\left[{\mathcal Q}_p\right]}, &&
S_n^{\text{total}} = \frac{\mathbb V_n^{\text{total}}\left[{\mathcal Q}_p\right]}{\mathbb V\left[{\mathcal Q}_p\right]}. \label{eq:sobol_indices}
\end{align}

\section{Application}
\label{sec:application}
We apply the \gls{uq} methods presented in the last section to two model problems. We first consider an academic example of an stochastic RLC circuit, since there is a closed-form solution available which allows us to illustrate the main ideas of the proposed approach in detail. We then consider the optical grating coupler \cite{preiner2008}, which is a non-trivial benchmark example from nanoplasmonics. 

\subsection{RLC circuit}
\begin{figure}
\begin{subfigure}[b]{.49\textwidth}
\centering
\includegraphics{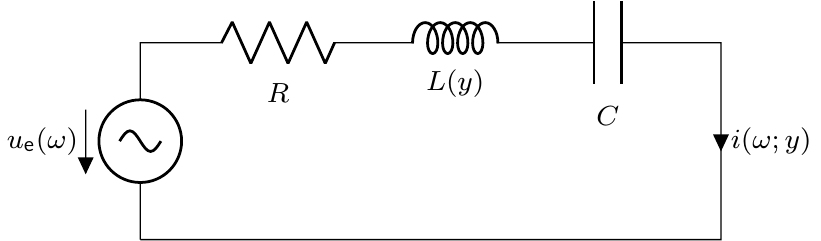}
\caption{Circuit diagram.}
\label{fig:RLC_circuit}
\end{subfigure}
\begin{subfigure}[b]{.49\textwidth}
\centering
\includegraphics{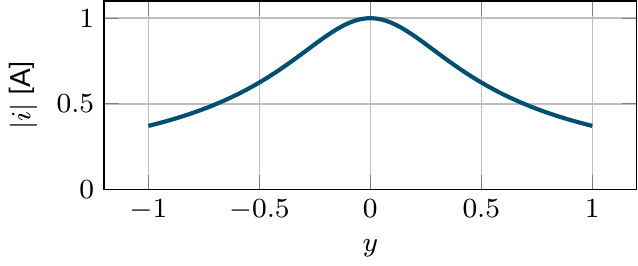}
\caption{Amplitude of electric current w.r.t. input parameter $y$.}
\label{fig:i_over_l}
\end{subfigure}
\caption{Benchmark problem: RLC circuit.}
\end{figure}
\begin{figure}
\begin{subfigure}[b]{.33\textwidth}
\includegraphics{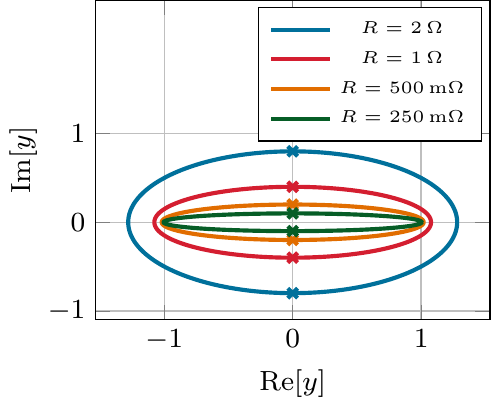}
\caption{Poles of $\mathcal Q(y)$ and corresponding Bernstein ellipses for different $R$.}
\label{fig:circuit_ellipses}
\end{subfigure}
\begin{subfigure}[b]{.33\textwidth}
\includegraphics{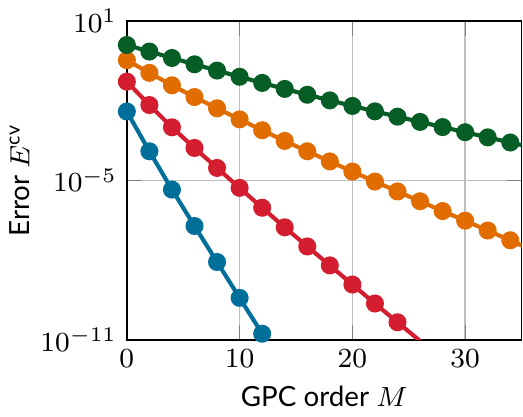}
\caption{Convergence of empirical $L^2$ error.}
\label{fig:Conv_gpc_l2}
\end{subfigure}
\begin{subfigure}[b]{.33\textwidth}
\includegraphics{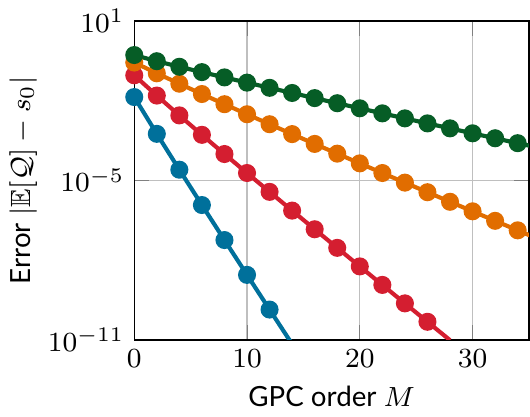}
\caption{Convergence of mean of GPC approximation.}
\label{fig:Conv_gpc_mean}
\end{subfigure}
\caption{Illustration of influence of poles in complex plane (RLC circuit).}
\end{figure}

We consider the model of an RLC circuit, as illustrated in Fig.~\ref{fig:RLC_circuit}. 
Assuming harmonic time dependency, the electric current $i$ is given by 
\begin{equation}
\Bigl(-L\omega^2 +j\omega R + \frac 1{C} \Bigr)i = j\omega  u_\text{e} \label{eq:rlc_ana_sol}
\end{equation}
We consider, arbitrarily chosen, an angular frequency $\omega=10^4\,$$\mathrm{s}^{-1},$ exciting voltage $u_\mathrm e = \SI{1}{V},$ capacitance  $C=\SI{10}{\micro\farad},$ and a (rather small) resistance of $R=\SI{1}{\ohm}$. Additionally, we consider a variable inductance $L(y)=\SI{1}{mH}+\SI{0.25}{mH}\cdot y$. The parameter $y$ is then modeled as a uniformly distributed random variable with probability density function $\rho=\mathcal U (-1,1)$, such that  a stochastic model is obtained. As \gls{qoi} $\mathcal Q$, we consider the amplitude of the current $\mathcal Q:=|i|$. 
Fig.~\ref{fig:i_over_l} shows the parametric dependency of the \gls{qoi} $|i|$ with respect to $y$, which is analytic for $y\in[-1,1]$. However, the continuation in the complex plane has poles at 
\begin{equation}
y =\pm i\frac R {\omega \cdot \SI{0.25}{mH}}. \label{eq:poles}
\end{equation}
This complex conjugate pole pair limits the size of the largest Bernstein ellipse, where $\mathcal Q(y)$ is analytic, which is illustrated in Fig.~\ref{fig:circuit_ellipses} for different values of $R$.

In each case, we compute \gls{gpc} approximations of increasing order for $\mathcal Q(y)$ using the \texttt{Chaospy} toolbox \cite{feinberg2015}. In particular, the \gls{gpc} coefficients of an $M-$th order approximations are computed by pseudo-spectral projection using Gaussian quadrature of order $M+1$. The accuracy of the surrogate models is then quantified in the empirical $L^2_\rho$ norm. In particular, we apply 
cross-validation using $N^\text{cv}=1000$ random parameter realizations $y_{\mathrm{cv}}^{(i)}$ drawn according to the probability density $\rho$, to compute the error 
\begin{equation}
E^\text{cv}= \frac {1}{N^\text{cv}} \sum_{i=1}^{N^\text{cv}} |\mathcal Q^\mathrm{PC}_M(y_{\mathrm{cv}}^{(i)})-\mathcal Q(y_{\mathrm{cv}}^{(i)})|^2. \label{eq:E_cv}
\end{equation}
Additionally, we compute the error in the first-stochastic moment, i.e. the mean value of the \gls{gpc} approximation given by the first polynomial coefficient $s_0$. The reference solutions for the expected values are obtained by Gaussian quadrature of order $200$ up to machine accuracy.  
The convergence of the corresponding surrogate model w.r.t. the polynomial order $M$, in terms of cross-validation and mean value accuracy, are presented in Fig.~\ref{fig:Conv_gpc_l2} and Fig.~\ref{fig:Conv_gpc_mean}, respectively. The plots confirm \eqref{eq:conv_est} numerically, showing a decreasing convergence order for decreasing values of $R$ corresponding to decreasing sizes of the associated Bernstein ellipses. Note that, according to \eqref{eq:poles} a similar behaviour as for decreasing damping can be expected for increasing amplitudes of the considered input variation.

\begin{figure}
\begin{center}
\begin{subfigure}{.43\textwidth}
\centering
\includegraphics{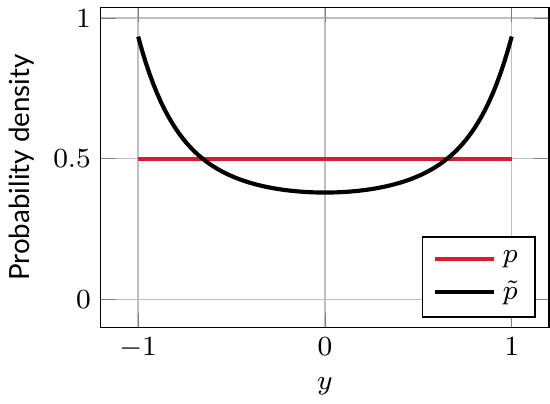}
\caption{Uniform input distribution and associated transformed density $\tilde p$.}
\label{fig:uniform_densities}
\end{subfigure}\hspace{1em}
\begin{subfigure}{.55\textwidth}\includegraphics{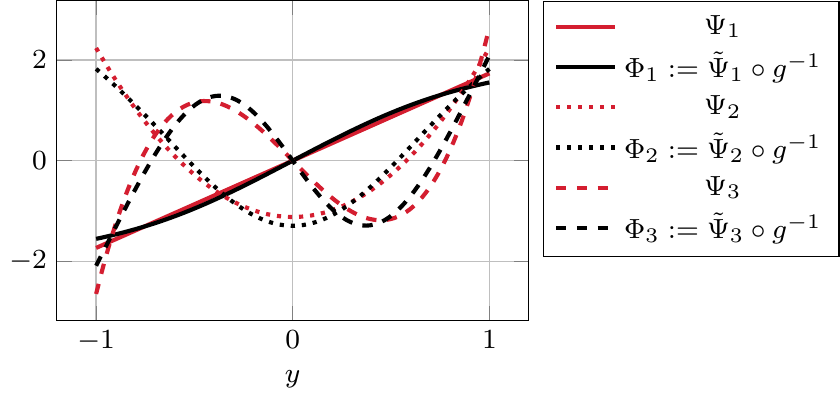}
\caption{Some basis functions for gPC and mapped gPC.}
\label{fig:uniform_basisfuns}
\end{subfigure}\vspace{2em}
\begin{subfigure}{.3\textwidth}
\centering 
\includegraphics{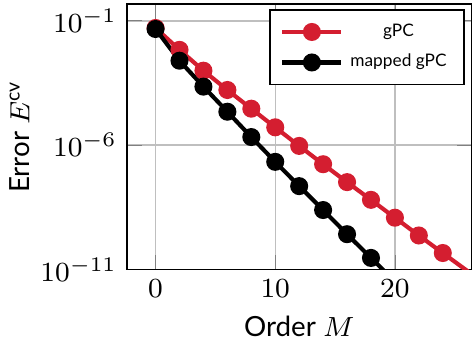}
\caption{Convergence of empirical $L^2_\rho$ error.}
\label{fig:uniform_cv_err}
\end{subfigure}\hspace{1em}
\begin{subfigure}{.3\textwidth}
\centering
\includegraphics{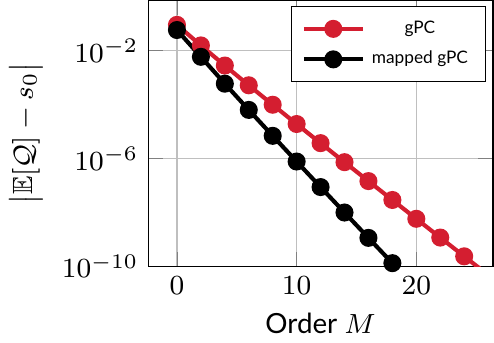}
\caption{Convergence of mean value.}
\label{fig:uniform_mean_err}
\end{subfigure}\hspace{1em}
\begin{subfigure}{.3\textwidth}
\centering
\includegraphics{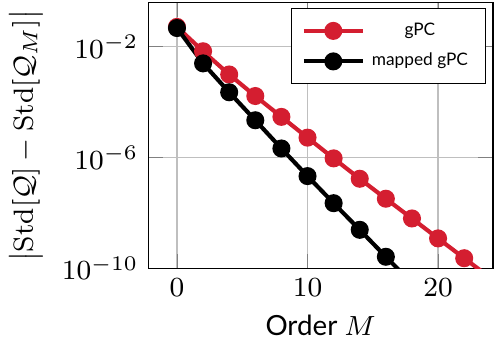}
\caption{Convergence of standard deviation.}
\label{fig:uniform_std_err}
\end{subfigure}
\end{center}
\caption{(Mapped) \gls{gpc} for stochastic RLC circuit with $R=\SI{1}{\ohm}$.}
\end{figure}

Next, we apply the conformally mapped \gls{gpc} expansions proposed in the last section. The implementation is done in \texttt{Python} based on \texttt{Chaospy} \cite{feinberg2015}. 
Fig.~\ref{fig:uniform_densities} shows the transformed density \eqref{eq:transformed_pdf} for a uniform input distribution $\rho$. Fig.~\ref{fig:uniform_basisfuns} depicts some exemplary basis functions of \gls{gpc} and mapped \gls{gpc}. Note that the \gls{gpc} basis functions are in this case Legendre polynomials, while the mapped basis functions, given by \eqref{eq:map_basis}, are no polynomials. 
We then study the convergence of the corresponding surrogate models, where mapped quadrature of order $M+1$ is used to compute the mapped \gls{gpc} expansions of order $M$. Fig.~\ref{fig:uniform_cv_err}, Fig.~\ref{fig:uniform_mean_err} and Fig.~\ref{fig:uniform_std_err} demonstrate the improved convergence order of the mapped approach, in terms of the cross-validation error, as well as the accuracy of the computed mean value and the computed standard deviation.

\subsection{Optical Grating Coupler}
\begin{figure}
\centering

\includegraphics{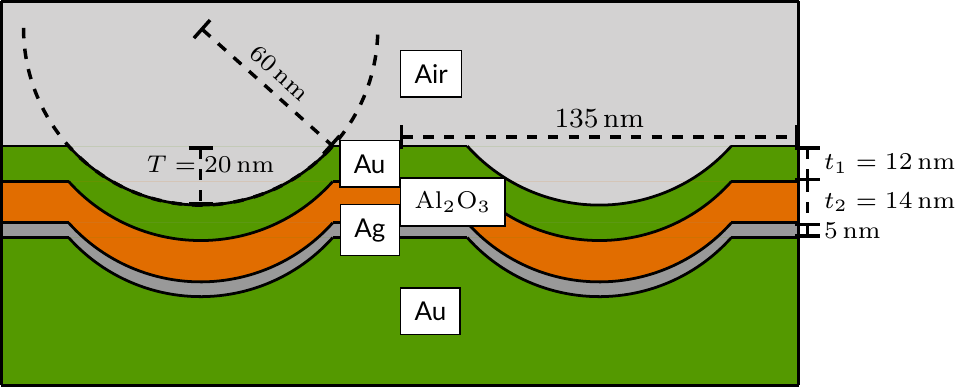}

\caption{Numerical model of considered optical grating coupler. Excitation by incident plane wave at upper boundary.}
\label{fig:design_coupler}
\end{figure}
We now consider the \gls{fe} model of an optical grating coupler \cite{preiner2008}, which was introduced in the beginning, see Fig.~\ref{fig:mimPlasmonMode}. 
The structure's design \cite{cstTutorial} is shown in Fig.~\ref{fig:design_coupler}. A plane wave at optical frequency hits the surface of the grating coupler. The incident wave couples with a \gls{mim} plasmon mode, which propagates along the metallic surface. It is found that the \gls{mim} resonance has a significant shift (in energy) as a function of the grating depth \cite{preiner2008} and therefore, it is of great interest to evaluate the influence of nano-technological manufacturing imperfections. 

\begin{figure}
\begin{center}
\begin{subfigure}{.43\textwidth}
	\centering
	\includegraphics{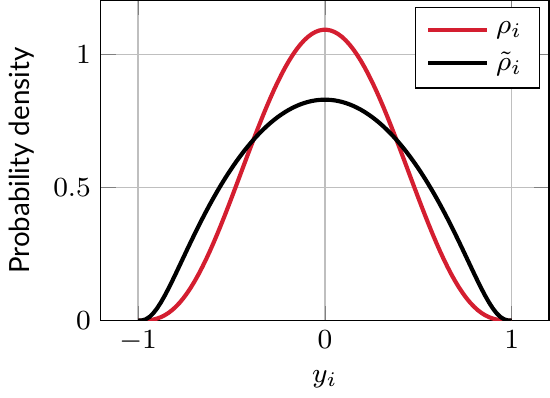}
	\caption{Beta input distribution $\rho_i$ and associated transformed density $\tilde \rho_i$.}
	\label{fig:beta_densities}
\end{subfigure}\hspace{1em}
\begin{subfigure}{.55\textwidth}
	\includegraphics{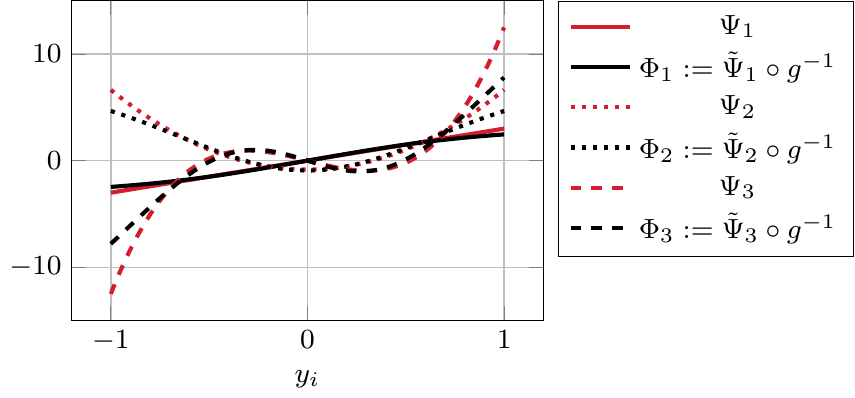}
	\caption{Some univariate basis functions for gPC and mapped gPC.}
	\label{fig:beta_basisfuns}
\end{subfigure}
\end{center}
\caption{\gls{gpc} for stochastic RLC circuit with beta distributed input parameter.}
\label{fig:beta_gpc}
\end{figure}

We use \textsc{FEniCS} \cite{alnaes2015} for the discretization and implement a design element approach \cite{braibant1984} for the geometry parametrization. The numerical model is described in greater detail in \cite{georg2018}. Note that we only consider periodic variations, modeling a systemic offset in the fabrication process, and do not address local uncertainties leading to different unit cells. Readers interested in the latter case are referred to \cite{schmitt2019optimization}.  The fundamental scattering parameter is considered as \gls{qoi} $\mathcal Q\in \mathbb C$. We consider three sensitive geometrical parameters as uncertain, in particular the thicknesses of the upper gold layer $t_1=\SI{12}{nm}+\Delta y_1$, the thickness of the dielectric layer $t_2=\SI{14}{nm}+\Delta y_2$ and the grating depth $T=\SI{20}{nm} + \Delta y_3$, as illustrated in Fig.~\ref{fig:design_coupler}. We model those parameters as independent beta distributed \glspl{rv} in the range of $\pm \Delta=\pm\SI{2}{nm}$. The corresponding shape parameters are chosen such that a normal approximation is approximated. The corresponding probability distribution $\rho_i$ of the \glspl{rv} $y_i,\,i=1,\ldots,3$ is shown in Fig.~\ref{fig:beta_densities}, together with the transformed density $\tilde \rho_i$. The univariate \gls{gpc} polynomials which are Jacobi polynomials in this case, as well as the mapped polynomials are illustrated in Fig.~\ref{fig:beta_basisfuns}.

\subsubsection{Decay of Fourier Coefficients}
We first study the decay of polynomial coefficients to numerically investigate the smoothness of the mapping from the input parameters to the complex S-parameter $\mathcal Q$ and justify the use of (mapped) polynomial approximations.
It has been shown, see e.g. \cite[Lemma 2]{nobile2009} where Legendre polynomials are considered, that if this mapping is smooth, the Fourier coefficients $s_\ensuremath{\mathbf{m}}$ of an $N-$variate \gls{gpc} approximation decay exponentially, i.e.
\begin{align}
|s_\ensuremath{\mathbf{m}}|^2 \le C e^{-\sum_{n=1}^N g_n m_n},
\end{align}
where $C$ and $g_n, ~n = 1,\ldots,N$ are positive constants independent of $\ensuremath{\mathbf{m}}$ and we have assumed that the polynomials are normalized.
We consider the maximum of the absolute value of the Fourier coefficients $s_\ensuremath{\mathbf{m}}$ with fixed maximum-degree $w = ||\ensuremath{\mathbf{m}}||_\infty := \max_i m_i$ 
\begin{align}
\max_{||\ensuremath{\mathbf{m}}||_\infty=w} |s_\ensuremath{\mathbf{m}}|^2 
\le \max_{||\ensuremath{\mathbf{m}}||_\infty=w} Ce^{-\sum_{n=1}^N g_n m_n} 
=  Ce^{-\min_{||\ensuremath{\mathbf{m}}||_\infty =w}\sum_{n=1}^N g_n m_n} 
\le Ce^{-(\min_n g_n) w}. 
\end{align}
It can be seen that the maximum Fourier coefficient is expected to decay exponentially with an increasing maximum-degree $w$.

\begin{figure}
\centering
\includegraphics{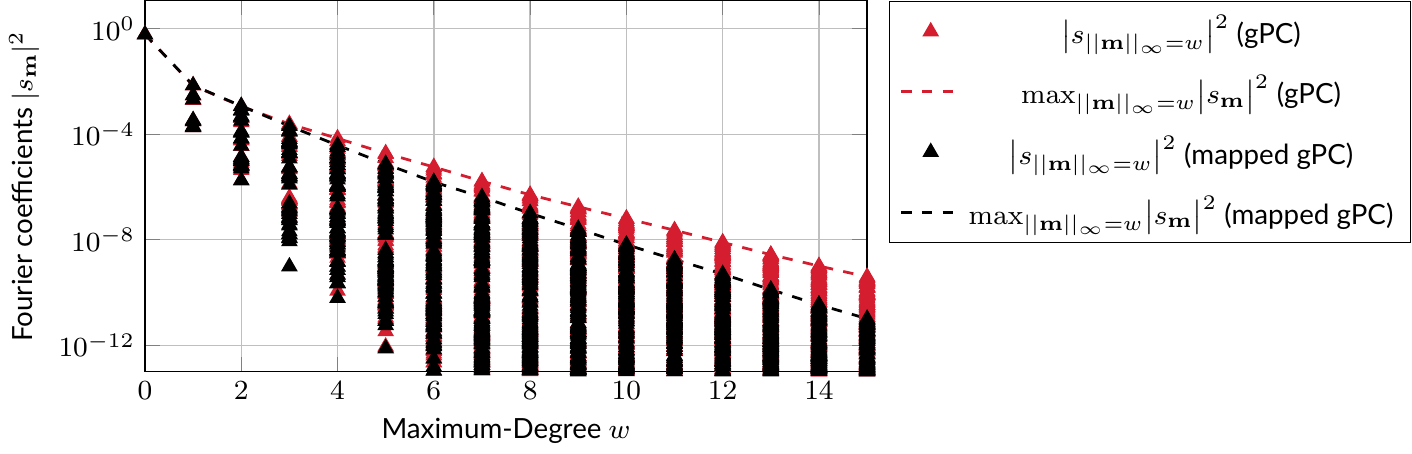}
\caption{Decay of Fourier coefficients of multivariate  (mapped) \gls{gpc} approximation.}
\label{fig:fouriercoeffs}
\end{figure}

We construct a multivariate \gls{gpc} approximation with a tensor-product basis of order $m_{\max} = 15$. The multivariate integrals of the pseudo-spectral projection are then computed by a Gauss quadrature of order 17.
All coefficients $s_\ensuremath{\mathbf{m}}$ are plotted in Fig.~\ref{fig:fouriercoeffs} in red color, where an exponential decay can indeed be observed.
This can be seen as a numerical indicator for smoothness of the approximated mapping $\mathcal Q(\ensuremath{\mathbf{y}})$. Additionally, we also construct a mapped approximation of same order and plot the corresponding coefficients in black color. It can be observed that the mapped coefficients exhibit a faster convergence and hence the mapped approach can be expected to show, again, an improved convergence. 
\subsubsection{Uncertainty Quantification}
\begin{figure}
\centering
\begin{subfigure}[b]{.3\textwidth}
\centering
\includegraphics{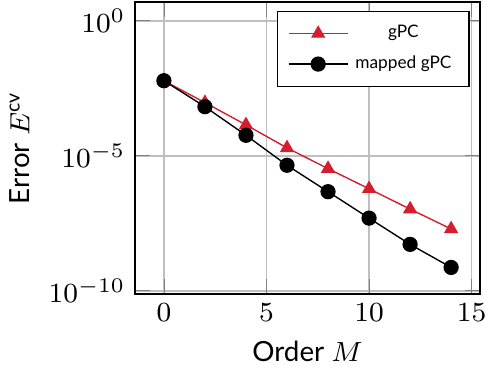}
\caption{Convergence of empirical $L^2_\rho$ error.}\label{fig:conv_coupler_l2}
\end{subfigure}\hspace{1em}
\begin{subfigure}[b]{.3\textwidth}
\centering
\includegraphics{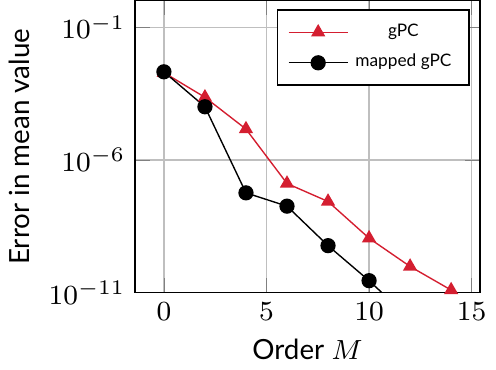}
\caption{Convergence of mean value.}
\label{fig:conv_coupler_mean}
\end{subfigure}\hspace{1em}
\begin{subfigure}[b]{.3\textwidth}
\centering
\includegraphics{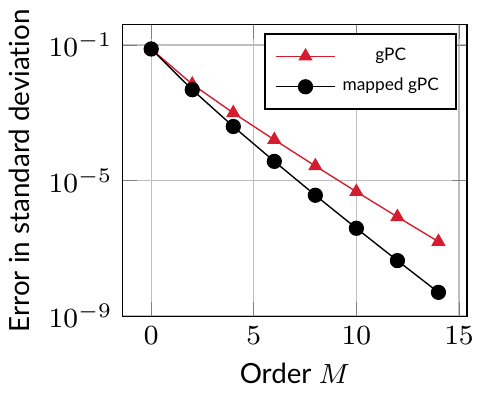}
\caption{Convergence of standard deviation.}
\label{fig:conv_coupler_std}
\end{subfigure}
\caption{Convergence of (mapped) \gls{gpc} expansions for the optical grating coupler.}
\end{figure}

Next, we consider approximations of the magnitude of the S-parameter $|\mathcal Q(\mathbf y)|$ using (mapped) tensor-product \gls{gpc} expansions of increasing order $M$, where pseudo-spectral projections of order $M+1$ is employed to compute the coefficients. 
Fig.~\ref{fig:conv_coupler_l2} compares \gls{gpc} and the proposed mapped counterpart in terms of the $L^2_\rho$-error \eqref{eq:E_cv}, in particular, again, by cross-validation with $10^3$ random parameter realizations. It can be observed that the mapped approach converges about $30\%$ faster w.r.t. the order $M$ than \gls{gpc}. However, the respective computational gain grows, in this case, exponentially w.r.t. the number of inputs and, hence, the required number of model evaluation to reach a prescribed accuracy reduces roughly by a factor of $2$. Similar findings hold for the stochastic moments, in particular, we present the convergence of the mean value in Fig.~\ref{fig:conv_coupler_mean} and the computed standard deviation in Fig.~\ref{fig:conv_coupler_std}. In this case, the reference solutions are obtained by Gaussian quadrature of order 30.

\begin{figure}
\centering
\includegraphics{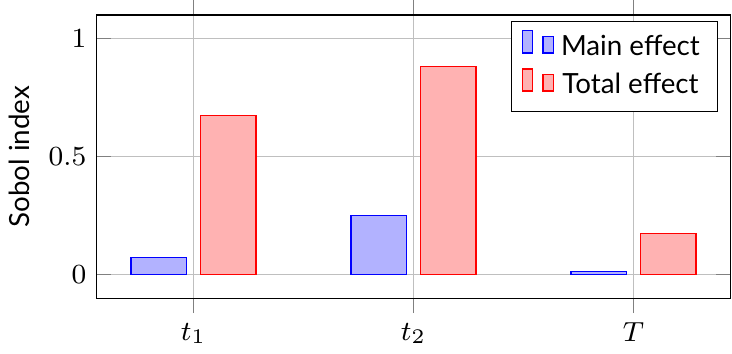}
\caption{Sensitivity of input parameters.}
\label{fig:coupler_sobol}
\end{figure}

Finally, the most accurate surrogate model, i.e. the mapped \gls{gpc} expansion of order $14$, is used to compute the mean value $\mathbb{E}[|\mathcal Q|]\approx0.786$ and the standard deviation $\sqrt{\mathbb{V}[|\mathcal Q|]}\approx 0.077$ of the \gls{qoi}. Additionally,  Sobol indices are computed and presented in Fig.~\ref{fig:coupler_sobol}. The thickness of the dielectric layer $t_2$ is identified as the most influential parameter. We note that there is a significant difference between the main- and total-effect indices. In particular, the sum of the first order indices is only $34\%$, while the remaining $66\%$ can be attributed to strong coupling effects among the parameters.

\section{Conclusions}
In this paper an efficient surrogate modeling technique for quantifying uncertainties in the material and geometry of high-frequency and optical devices was presented. The proposed method is based on \gls{gpc} to achieve spectral convergence. Through a combination with conformal maps we were able to enlarge the region of analyticity. This lead to an improved convergence rate, which was numerically demonstrated for two benchmark problems. In particular, the approach showed significant gains in either accuracy or computational cost, without adding any relevant extra computational effort. Due to orthogonality of the proposed basis, stochastic moments as well as Sobol indices can be directly obtained from the coefficients. It is worth noting that this technique can also be combined with other techniques for convergence acceleration such as adjoint-error correction, sparse-grids and (adjoint-based) adaptivity for the multivariate case \cite{georg2018}.

\subsection*{Acknowledgements} 
This research was funded by the Deutsche Forschungsgemeinschaft (DFG, German Research Foundation) -- RO4937/1-1. The work of Niklas Georg is also partially funded by the \emph{Excellence Initiative} of the German Federal and State Governments and the Graduate School of Computational Engineering at Technische Universit\"at Darmstadt.

\end{document}